\documentclass[12pt]{iopart}
\begin{document}
\title[Magnetotransport in an array of magnetic antidots]
{Magnetotransport in an array of magnetic antidots}
\author{L. Solimany}
\address{Center for Theoretical Physics and Mathematics,\\
P.O.Box 11365-8486,
 AEOI,Teheran-Iran\\}
\begin{abstract}
Classical transport properties of an electron, moving in plain,
in an array of magnetic antidots has been calculated.
Homogeneous magnetic field $\vec B = B \vec e_z$ fills the whole
space except of cylinders of radius 
$r_0$, which are arranged in a square lattice with a lattice
constant $a$.
The magnetoresistance shows additional peaks and minima by varying the strength
of the magnetic field, according to pinned
orbits at antidots and to propagating orbits in transport direction,
respectively.

\end{abstract}
\maketitle

The dynamics of charged particles in spatially
modulated magnetic field give rise to a variety of
interesting phenomena.
The motion of ballistic electrons in modulated magnetic
field is also believed to be closely related to the motion of
composite fermions in a density modulated 2DEG in the
 fractional quantum Hall regime~\cite{FQHE1,FQHE2}.
 Commensurability effects
 in  weak periodic magnetic fields were predicted~\cite{PMF1,PMF2} and
 observed for modulation in one~\cite{Weis1} and two
 directions~\cite{Weis2} experimentally. Such periodic magnetic
 field can be generated by using either patterned
 superconducting~\cite{Carmona} or ferromagnetic gates
 ~\cite{Weis1},\cite{Izawa}.
Closely related to our system are antidot arrays,
which consist of periodically arranged voids in an electron gas.
Charged particles move in this two dimensional potential landscape
under the influence of perpendicular magnetic field and
are scattered on the antidots.~\cite{antidot}
Additional peaks in the low field magnetoresistance have been
observed~\cite{Weis3} in these antidot arrays and
explained~\cite{Fleischmann} by a pinning mechanism of classical
circular orbits
enclosing 1,2,4,9 and 21 antidots in terms of nonlinear dynamics.
The nature of magnetoresistance peaks are caused by islands of
regular motion due to nonlinear resonances.\\
In the present paper we study the magnetotransport in an array
of {\it magnetic antidots}. Instead of antidot scatterers we introduce
circular regions into the system with zero magnetic field inside. Inside
the magnetic antidots the electron moves in a field-free region
and outside is its motion described by the Lorentz force. As demostrated
in Fig.~\ref{fig:model} the inhomogenity of the z-componetnt of the
magnetic field for one magnetic antidot is given by
$$
B(r)\vec e_z=\left\{
\begin{array}{ccc}
\displaystyle{-B}\vec e_z & \mbox{for} &  r>r_0 \\
0                          & \mbox{else} &     
\end{array}\right.
$$
and these antidots are arranged in a rectangular array
of lattice constant $a$.
In the recent experiments on modulated magnetic fields,
the period of modulation lies by 700 nm and is larger then
the Fermi wave length, but smaller then the elastic mean free
path (10 $\mu$m) and thus the dynamics of the wave packet approaches
the classical limit. Magnetic antidot array could be realized using
supeconducting
materials, which  would be maped on two dimensional electron gas
in a rectangular array arrangement. By applying magnetic field,
in circular region of superconducting material, the applied magnetic field
would be swept out and the electrons would move in a system described in this
paper. In the calculation we will use a system, that would be produced
by an ideal vortex with a constant magnetic field outside and zero
magnetic field inside, instead of more realistic exponential distribution
of the magnetic. We consider also one elctron approximation.\\
The classical trajectory is found as the solution to Newtons' equation
of motion with the force given by the Lorentz expression
$\vec F=e\vec v\times \vec B$ for a particle of charge $e$. It consist,
as well known of straight line segments inside the antidot
and an arc of a circle outside, with the radius of the curvature given
by the cyclotron radius $r_c=\frac{v}{\omega_c}$ with $v$ being
the particle velocity and $\omega_c=\frac{eB}{m}$ the cyclotron
frequency. There exist also undisturbed circular orbits, which do not
intersect any antidot.
The classical approximation for the dynamics of an electron wave
packet in a modulated magnetic field is described by the Hamiltonian
for one magnetic antidot
$$
H(x,y,p_x,p_y)=\left\{
\begin{array}{ccc}
[(p_x+eBy/2)^2+(p_y-eBx/2)^2]/2m & \mbox{for} &  r>r_0 \\
\vec p ^2/2m                          & \mbox{else} &     
\end{array}\right.
$$
where $m$ is the effective mass
of the electron.
First we introduce some characteristic parameters of the system
and express all quantities in dimensionless
units: the coordinate $x\prime=x/a$, $y\prime=y/a$,$t\prime=t/\tau_0$
and $H\prime=H/\epsilon_F$ with $\epsilon_F$, the Fermi energy,
$\tau_0=(\epsilon_F/ma^2)^{1/2}$ . The magnetic field is scaled by
$B_0$ at which the cyclotron radius corresponds to $a/2$.
Calculating the Poisson brackets and omitting
the primes
we obtain following equations of motion for one magnetic antidot
$$\dot x=v_x,\ \ \ \dot y=v_y$$
$$
\dot v_x=\left\{
\begin{array}{ccc}
B/B_0v_y & \mbox{for} &  r>r_0 \\
0                       & \mbox{else} &     
\end{array}\right.
$$
$$
\dot v_y=\left\{
\begin{array}{ccc}
-B/B_0v_x & \mbox{for} &  r>r_0 \\
0                       & \mbox{else} &     
\end{array}\right.
$$
Choosing $r_0=0.3a$ and using a numerical integration method
it is possible to calculate
the trajectories of the electron for different values of magnetic field.
We investigate the motion in phase space ($x,y,v_x,v_y$) by means
of Poincar\'e surface of section at ($x = 0\ \mbox{mod}\ a$). Since
the energy $E=v_x^2/2m+v_y^2/2m$ of the system is conserved,
we have to consider
the surface of section ($y,v_y$) for $v_x>0$ and $v_x<0$ separately.
The initial conditions ($y_i,v_{yi}$) lead for $v_x>0$ and $v_x<0$
to different trajectories and the corresponding two
surfaces of section have mirror symmetry.
Fig.\ref{fig:SOS} shows the surface of section for $v_x>0$ for three different
values of magnetic field $B/B_0=0.5,0.6\ \mbox{and}\  1.0$, the corresponding
cyclotron radius is $r_c= a,\ 0.83a,\ a/2$ .
The whole phase space
consits at $B/B_0=0.5$ of chaotic sea as shown in Fig.\ref{fig:SOS} (c),
 one of possible chaotic trajectories
in coordinate space is shown in
Fig.\ref{fig:traj} (IV).
At  $B/B_0=0.6$ we found  periodic orbits, which intersect
two magnetic antidots as shown in Fig.\ref{fig:traj} (III).
These orbits occupy a relative small portion in the phase space
as shown in  Fig.\ref{fig:SOS} (b),
 label(III). 
An other type of periodic orbits, which intersect four magnetic antidots
is shown in Fig.\ref{fig:traj} (V).  The fingerprint of this
orbit in the surface of
section is shown in Fig.\ref{fig:SOS} (a), label (V).
 When the cyclotron radius approaches half of the
lattice constant, a different type of periodic orbit becomes dominat.
These orbits are pinned around one magnetic antidot as shown in
Fig.\ref{fig:traj} trajectory (II)
 and build in 
the surface of section a family of eliptic fixed points in a circular region
at $(y,v_y)=(0.5,0)$ of diameter $r_0$ as shown in Fig.\ref{fig:SOS} (a),
region (II).
These orbits can also
intersect one magnetic
antidot for higher magnetic fileds as shown in Fig.\ref{fig:traj} (I) and
dominate the dynamics of the system.
The orbits of type (I),(II) and (V) are present in a wide sector
of magnetic field strength $B/B_0 \in ]0.8,1.7]$.
For lower values of magnetic field
$B/B_0=0.3$
periodic orbits, which intersect eight magnetic antidots occupy
a relative small region (10\%) in phase space.
For $B/B_0>2.3$ all possible orbits are pinned, and no classical
chaotic trajectories exist anymore. The inset of Fig.\ref{fig:rxx} (a)
 shows the portion $p$
of regular orbits in dependence of magnetic field.\\
At low temperatures elastic impurity scatterings with mean
scattering time $\tau$ have to be considered.
Using the classical Kubo formula~\cite{Fleischmann} which
includes these scattering, we can
calculate the conductivity $\sigma_{ij}$ in dependence of magnetic field.
$$\sigma_{ij}(0,\tau)=\frac{ne^2}{k_BT}\int_0^\infty dt e^{-1/\tau}
<v_i(t)v_j(0)>_{\Gamma_c}$$
 $<v_i(t)v_j(0)>$ is the
velocity correlation function averaged over chaotic part of the
phase space $\Gamma_c$,
$k_B$ is the Boltzmann constant, $T$ temperature and $n$ the
2D electron density. The magnetoresistance can be obtain by inversion
of the conductivity tensor
$R_{xx}=\sigma_{xx}/(\sigma_{xx}^2+\sigma_{xy}^2)$ and
$R_{xy}=\sigma_{xy}/(\sigma_{xx}^2+\sigma_{xy}^2)$.
We have calculated the magnetoresistance $R_{xx}$ and the Hall resistance
$R_{xy}$ in dependence of magnetic field for a typical
value of $\tau/\tau_0=30$.
The magnetoresistance shown in Fig.\ref{fig:rxx} (a) displays a
dominant peak at $B/B_0\approx 1$ which corresponds to pinned
orbits around one antidot. According to the portion of
pinned orbits, see the inset of Fig.\ref{fig:rxx} (a) we conclude,
that this peak is caused by these periodic orbits.
We should expect also a peak at $B/B_0\approx  0.3\  \mbox{and}\  0.6$
as in the case of antidot lattices, but instead we observe a minimum
at $B/B_0=0.8$.
The main difference to antidot lattices, where electrons are scattered
on antidots, is that the sign
of the angular momentum of an electron, moving in a magnetic antidot array
remains conserved.
The magnetoresistance is not
only caused by islands of regular motion in the phase space
(pinned orbits), but also
by orbits, which propagate in the direction
of transport. For $B/B_0=0.8$ we found orbits, which propagate in
transport direction, as shows in Fig.~\ref{fig:traj} (b) and explain
the minimun in the magnetoresistance at this strength of
magnetic field.                 
We obtain even $R_{xx}/R_{xx}(0)<1$, the resistance lies lower
then, for the case of zero magnetic field.
The Hall resistance as shown in Fig.~\ref{fig:rxx} (b) shows
steps at magnetic field values for which the magnetoresistance $R_{xx}$
lies in minimum.\\
In conclusion, classical transport properties of electrons,
moving in a magnetic antidot array has been explained by pinned
orbits at antidots and by propagating orbits in transport direction.\\
I thank Prof.B.Kramer for the initial ideas of this work.

\newpage
{\bf Figure captions}\\

\begin{figure}[h]
\caption{Schematic model of the magnetic antidot array. Inside the
antidot is fieldfree region and outside acts homogeneous magnetic field in
$z$-direction.The radius of the antidot is $r_0$ and the lattice
constant is $a$.}
\label{fig:model}
\end{figure}
\begin{figure}[h]
\caption{The Poincar\'e surface of section for three differnt
values of magnetic field strength.
 (a) $B/B_0=1.0$, a circular region at $(y,v_y)=(0.5,0)$ is filled
 by elliptic fix points, which correspond to pinned orbit of type II.
 Pinned orbits of type V are still present.
(b) $B/B_0=0.6$, islands of regular motion
 corresponding to pinned orbits of type III and V are present.
(c)$B/B_0=0.5$, the chaotic sea fills
the whole phase space. }
\label{fig:SOS}
\end{figure}
\begin{figure}[h]
\caption{Different types of trajectories in magnetic antidot array.
(a) Pinned orbits, which cause peaks in magnetoresistance.
(b) Propagating orbits in transport direction, which cause
minima in magnetoresistance.}
\label{fig:traj}
\end{figure}

\begin{figure}[h]
\caption{ The magnetoresistance (a) and Hall resistance (b) in dependence
of the strength of the magnetic field. The minimum at $B/B_0=0.8$
 in $R_{xx}$ is caused
by propagating orbits in transport direction and the the peak at $B/B_0\approx
1.0$ is caused by pinned orbits around one magnetic antidot.
The inset shows the amount of islands of regular motion in phase space.
Steps in $R_{xy}$ correspond to negative gradients in $R_{xx}$.}
\label{fig:rxx}
\end{figure}

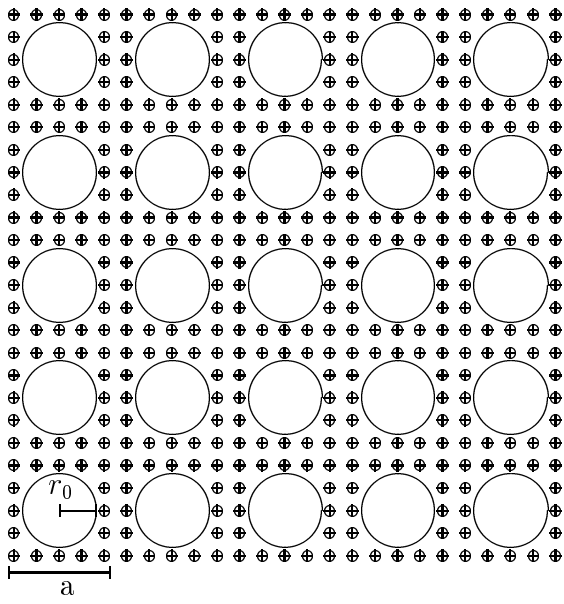
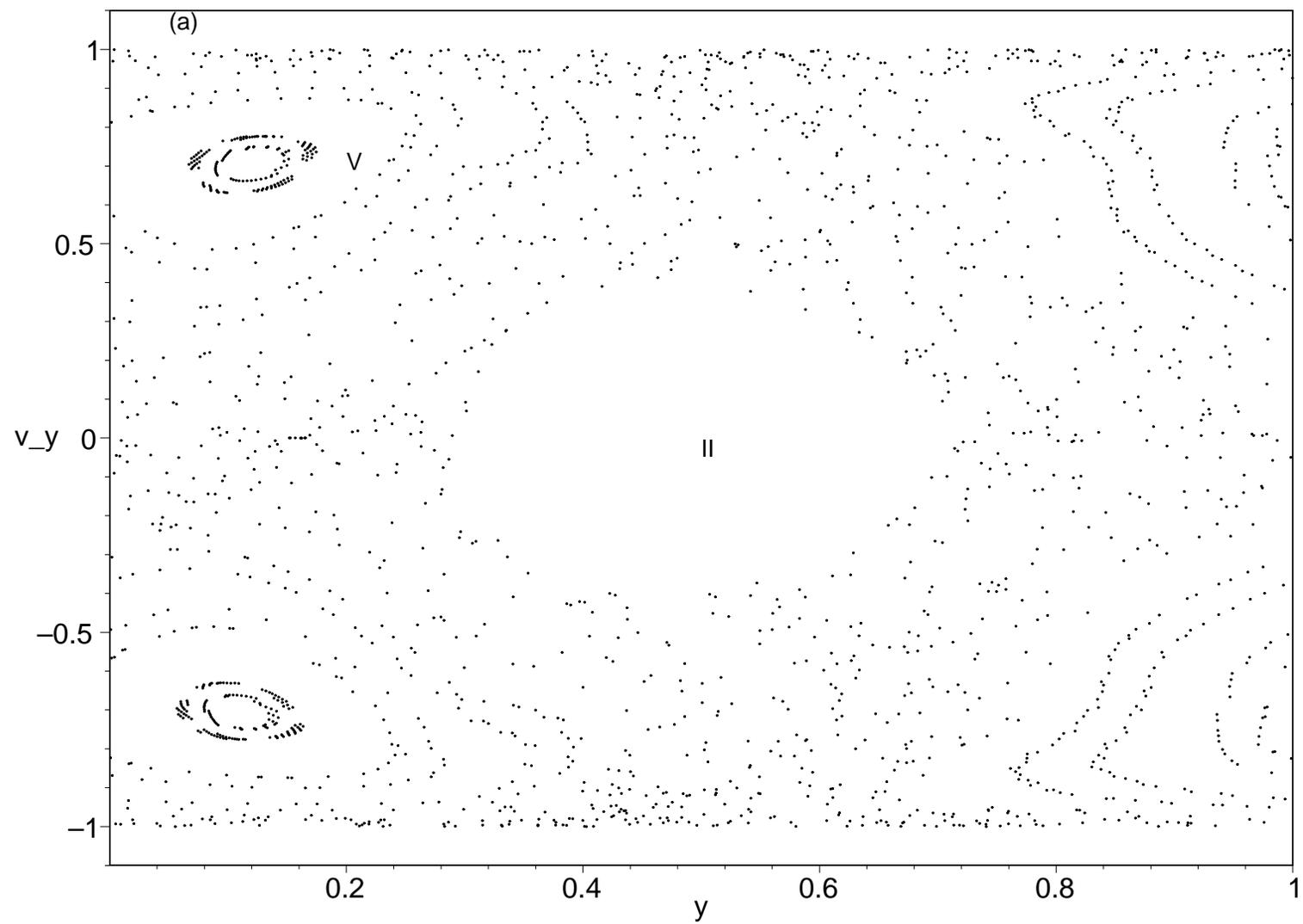
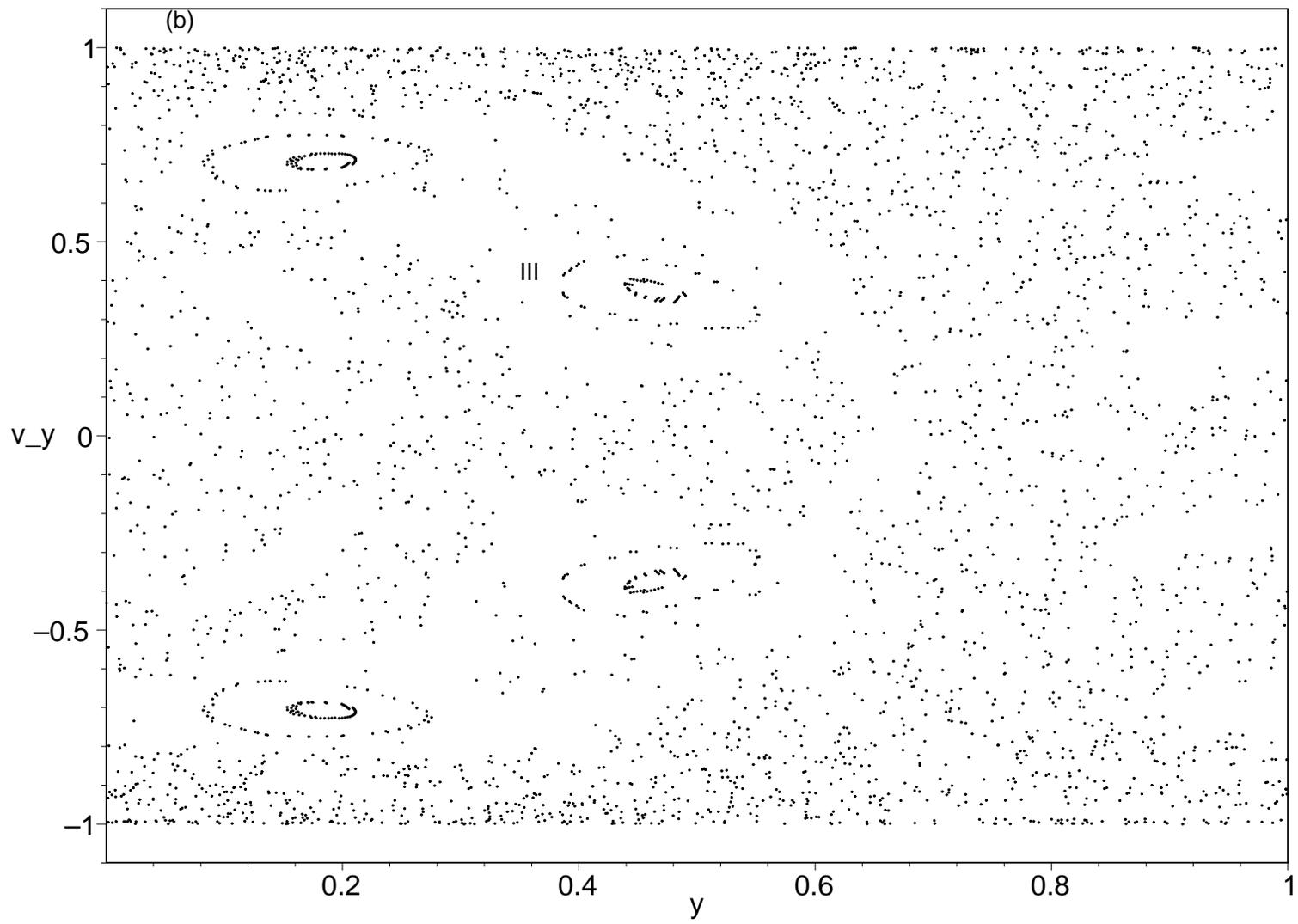
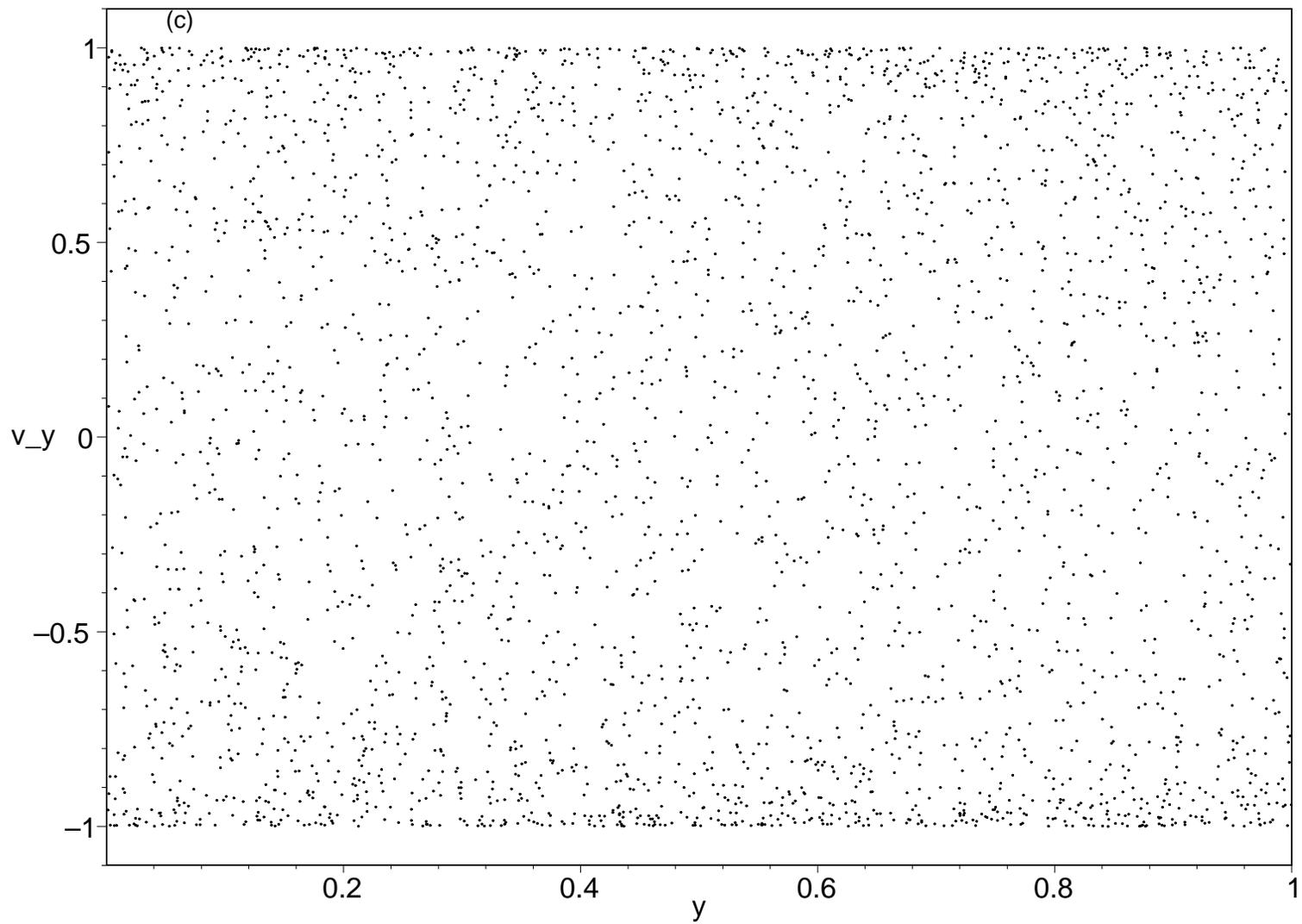
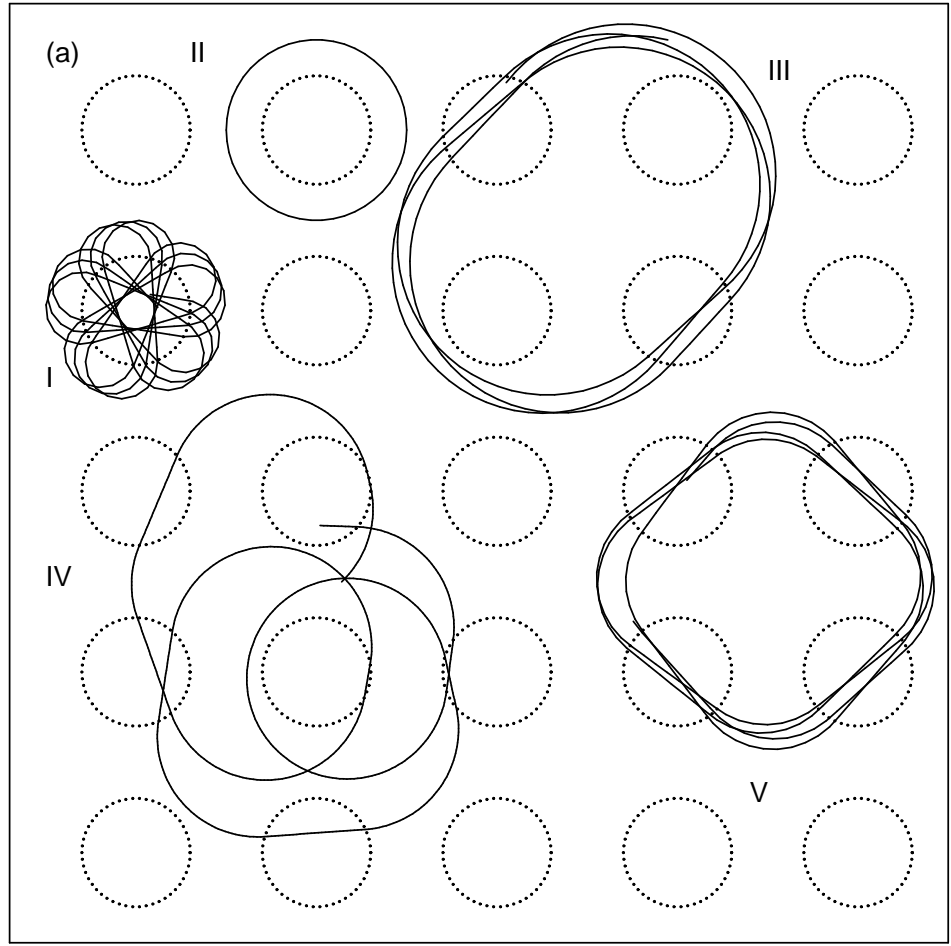
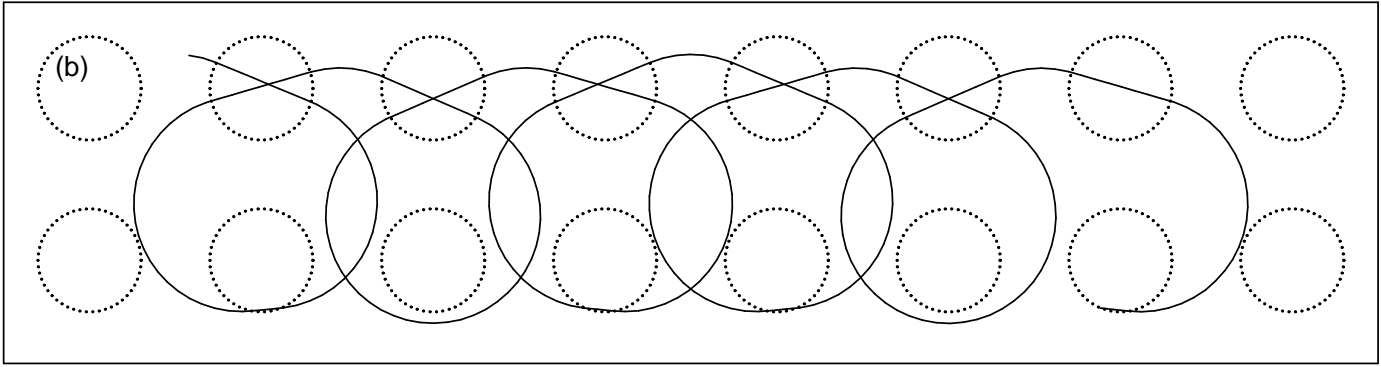
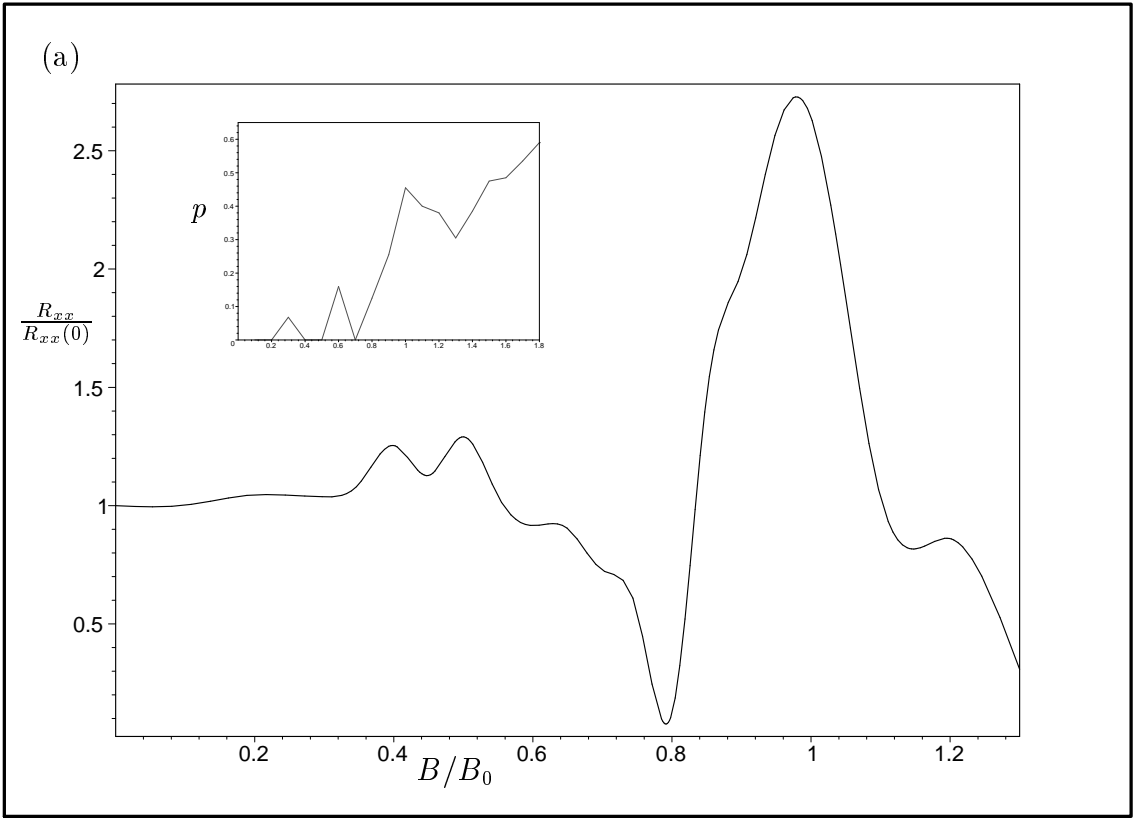
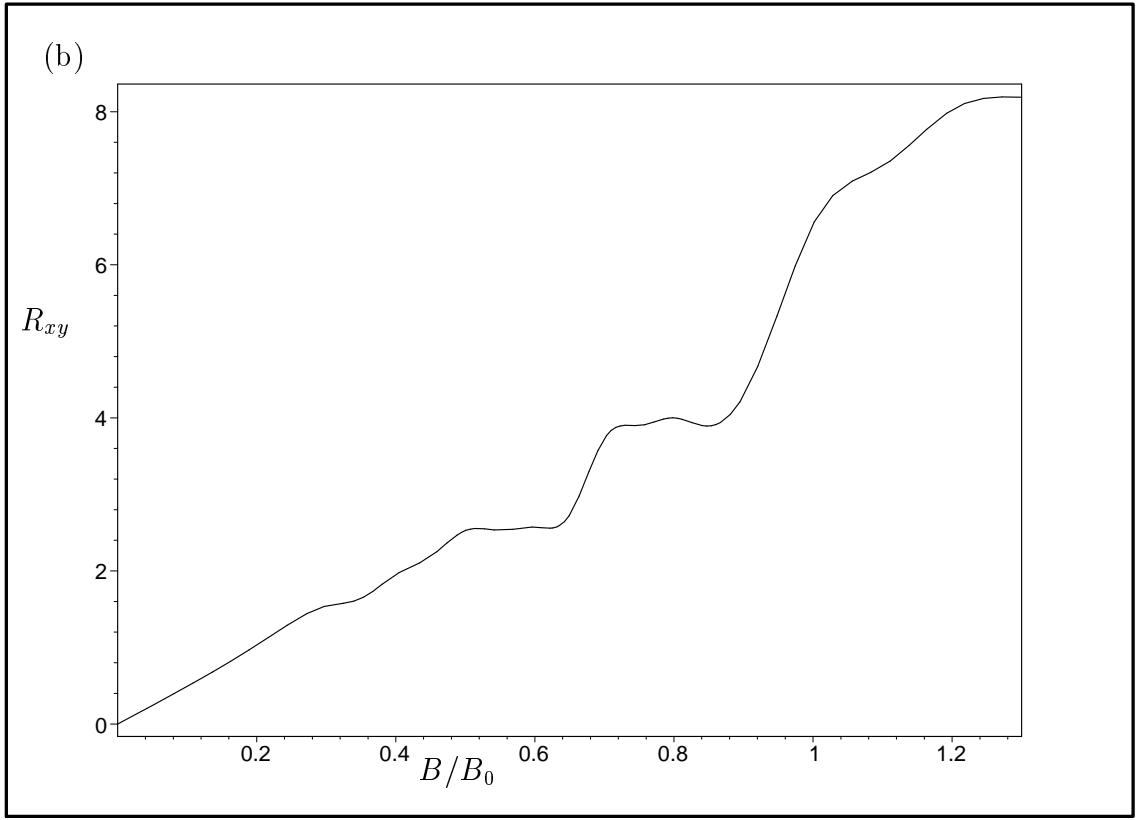
\end{document}